# *The Information-Theoretic View of Quantum Mechanics and the Measurement Problem(s)*

Federico Laudisa

Department of Philosophy and Humanities, University of Trento
Via Tommaso Gar 14, 38122, Trento, Italy

**Abstract**

Until recently Jeffrey Bub and Itamar Pitowsky, in the framework of an information-theoretic view of quantum mechanics, claimed first that to the measurement problem in its ordinary formulation there correspond in effect *two* measurement problems (simply called the *big* and the *small* measurement problems), with a different degree of relevance and, second, that the analysis of a quantum measurement is a problem only if other assumptions – taken by Pitowsky and Bub to be unnecessary 'dogmas' – are assumed. Here I critically discuss this unconventional stance on the measurement problem and argue that the Bub-Pitowsky arguments are inconclusive, mainly because they rely on an unwarranted extension to the quantum realm of a distinction concerning the foundations of special relativity which is in itself rather controversial.

1. **Introduction**
2. **How many measurement problems are there?**
3. **A 'kinematical' solution to the measurement problem?**
4. **The Bub-Pitowsky (dis)solution does not work**
5. **The *In principle underdetermination* claim**
6. **Conclusions**

## 1. Introduction

In retrospect, it is far from surprising that in one of the great scientific works of the XXth century, the 1932 von Neumann book on the mathematical foundations of quantum mechanics (QM), an entire chapter is devoted to the problem of how to construct an ideal quantum-mechanical model of a measurement (von Neumann 1955, chapter VI). The von Neumann treatment, and the place occupied by this problem in his first formally rigorous formulation of quantum theory, already revealed how controversial the status of measurement in QM would have been, to the extent that the very notion of measurement would turn out to be the *locus classicus* for emphasizing the lack of consensus on the interpretation of the theory. In fact, that chapter happens to be the major source of what is usually defined as the measurement *problem*: unlike the case of the vaguely defined classical-quantum distinction previously advocated by Bohr, von Neumann explicitly confronts the implications of the assumption that – in the context of a measurement of a physical quantity on a quantum system *S* with an apparatus *A* – the laws of QM govern *both S and A*. As a matter of fact, the measurement problem is widely taken to be a true touchstone for a classification of the main different interpretations of QM. If, according to the folklore, a standard measurement procedure in QM induces a 'collapse' of a superposition state of the joint system *A+S* into one of its components, the disagreement arises at the starting block: is this 'collapse' a real physical process or not? The No-Answer leads to the claims that such 'collapse' is just a sort of perspective effect (Everett-style of thinking), or a phenomenological effect due to the inaccessibility of some physical variable relevant to the measured system (Bohmian-style of thinking). On the other hand, the Yes-Answer leads to the assumption of new laws to be added to standard QM, laws that dictate the how and when of a physical collapse, so as to make it compatible with the empirical fact that at the end of a measurement we obtain a definite outcome (GRW-style of thinking)[1]. Independently of the problem of how the collapse of the post-measurement state of *A+S* is to be interpreted, though, there are those who take a different approach and question the very idea of a *real* measurement problem in its ordinary formulation. In this approach, developed in more recent times especially by the late Pitowsky and

---

[1] As is well known, the number of interpretation largely exceeds the list including Everett, Bohm and GRW, but I take these to be the most transparent in taking an unambiguous stance on the issue of the exact nature of collapse. Moreover, also Everettians and Bohmians can be said to 'add' something to the QM description, although in a more nuanced way, given that their view are predictively equivalent to standard QM: the Everettians add an explanation of why we perceive just *one* world, whereas the Bohmians add information on the position of the quantum systems under scrutiny, information that is supposed to be unavailable in the standard formulation.

Jeffrey Bub on the background of the so called 'information-theoretic' view of QM, it is claimed first that to the measurement problem in its ordinary formulation there correspond in effect *two* measurement problems (simply called the *big* and the *small* measurement problems), with a different degree of relevance and, second, that the analysis of a quantum measurement is a problem only if other assumptions – taken by Pitowsky and Bub to be unnecessary 'dogmas' – are assumed.

In the present paper I will try to critically assess this unconventional stance on the measurement problem. In section 2, I will investigate the very definition of the big and the small measurement problems provided by Pitowsky and Bub, in order to clarify their logical status and mutual relationships. Since Pitowsky and Bub take the status of these problems *qua* problems to depend on two claims that, in turn, they take to be true dogmas of the folklore view of QM, an integral part of my analysis will be to clarify how the big/small distinction fares with respect to these two 'dogmas'. In section 3 I will review how what is called the 'big' measurement problem is claimed to be (dis)solved in the information-theoretic view of QM to which Bub and Pitowsky subscribed. In section 4 I will argue that their (dis)solution proposal does not work, since it relies on an unwarranted extension to the quantum realm of a stance concerning the foundations of special relativity, whereas in section 5 I will deal with a further aspect of the Bub-Pitowsky (dis)solution proposal, connected with the extent to which alternative interpretations of QM should be accepted even if they do not provide new empirical predictions. My overall conclusions will be drawn in the final section 6.

## 2. How many measurement problems are there?

Jeffrey Bub and Itamar Pitowsky initially present the measurement problem in a (rather) standard way:

> The measurement problem is the problem of explaining the apparently 'irreducible and uncontrollable disturbance' in a quantum measurement process, the 'collapse' of the wavefunction described by von Neumann's projection postulate." (Bub, Pitowsky 2010, p. 438).

As we will see later, Bub and Pitowsky support an information-theoretic approach to QM but with a realistic tone, and as a consequence they are unhappy with a purely instrumentalistic reading of the collapse. In this vein, they claim that the 'irreducible and uncontrollable measurement disturbance' – vaguely associated with collapse in a Copenhagenish style – fails to receive a decent explanation:

> Without a dynamical explanation of this measurement disturbance, or an analysis of what is involved in a quantum measurement process that addresses the issue […], the theory qualifies as an algorithm for predicting the probabilities of measurement outcomes, but cannot be regarded as providing a realist account, in principle, of how events come about in a measurement process. (Bub, Pitowsky 2010, p. 435).

This statement might be taken simply as a rather common manifestation of dissatisfaction with instrumentalism in the interpretational debate on QM, were it not for the *dynamical* qualification of the explanation that Bub and Pitowsky take to be necessary. This reference, to which we will come back later, is also explicit in the claim that there is no just one measurement problem, but in fact *two*:

> The 'big' measurement problem is the problem of explaining how measurements can have definite outcomes, given the unitary dynamics of the theory: it is the problem of explaining *how individual measurement outcomes come about dynamically.* The 'small' measurement problem is the problem of accounting for our familiar experience of a classical or Boolean macroworld, given the non- Boolean character of the underlying quantum event space: it is the problem of explaining the *dynamical emergence of an effectively classical probability space of macroscopic measurement outcomes* in a quantum measurement process. (Bub, Pitowsky 2010, p. 438, my emphasis)[2]

As is clear from the semantics of the 'big/small' distinction, the small measurement problem is taken to be relatively easy to solve. Putting aside for a moment the issue of characterizing different layers of the natural world as 'Boolean' or 'non-Boolean' (a far-from-innocent issue, to which we will return later in connection with the 'big' measurement problem), Bub and Pitowsky do not diverge from the mainstream approach where decoherence does the job:

> The 'small' measurement problem is resolved by considering the dynamics of the measurement process and the role of decoherence in the emergence of an effectively classical probability space of macroevents to which the Born probabilities refer." (Bub, Pitowsky 2010, p. 438)[3].

---

[2] Brukner 2017 also proposes a view according to which the measurement problem, as ordinarily understood, in fact splits into two versions. The Brukner overall approach to the solution to the measurement(s) problem, however, has a 'relational' flavour and significantly diverges from the Bub-Pitowsky approach.

The heart of the matter lies in the 'big' measurement problem, whose status is controversial according to Bub and Pitowsky:

> The big measurement problem depends for its legitimacy on the acceptance of two dogmas. [...] The first dogma is Bell's assertion that measurement should never be introduced as a primitive process in a fundamental mechanical theory like classical or quantum mechanics, but should always be open to a complete analysis, in principle, of how the individual outcomes come about dynamically. The second dogma is the view that the quantum state has an ontological significance analogous to the ontological significance of the classical state as the 'truthmaker' for propositions about the occurrence and non-occurrence of events, i.e., that the quantum state is a representation of physical reality." (Bub, Pitowsky 2010, p. 438, my emphasis).

In the present section, I will take for granted the Bub-Pitowsky formulation of the big measurement problem and I will consider its logical relation with the two dogmas, independently from the details of that formulation, whereas I will focus on the controversial status of the formulation itself in the next section.

The alleged dependence of the big measurement problem on the first 'dogma' is unconvincing for two main reasons. First, the rejection of Bell's meta-theoretical stance, and the resulting inclusion of measurement among the primitive theoretical notions, are far from explaining away the big measurement problem *per se*. The vaguely defined coexistence of unitary and non-unitary dynamics has been considered puzzling and unsatisfactory since the origins of QM, *quite independently* from whether measurement should have a primitive or derivative status in a fundamental theory such as QM. Therefore, even if we decide to drop the first 'dogma', this does not put us in any better position to solve, or *dis*solve, the big measurement problem. Second, the very fact that the problem of coexistence of unitary and non-unitary dynamics, recalled above, has been widely acknowledged as a crucial issue – if not *the* issue – in the foundations of QM should suggest caution in qualifying the Bell meta-theoretical stance as a 'dogma': it is a legitimate assumption that, as it happens to many other options on the market, comes with its possible strengths and weaknesses. To turn it into a 'dogma' is no argument, on the contrary it is an excercise in dogmatism itself.

---

[3] It is widely believed that also this solution is a for-all-practical-purposes solution, at least in the standard presentation in which there is just one measurement problem (for a recent review Bacciagaluppi 2020).

The dependence of the big measurement problem on the second dogma seems even more problematic. In an earlier paper, Pitowsky had described the second dogma as the assumption according to which

> the quantum state is a real physical entity, and that denying its reality turns quantum theory into a mere instrument for predictions. *This last assumption runs very quickly into the measurement problem*." (Pitowsky 2006, p. 214, my emphasis).

Since, according to Pitowsky, "the BIG problem concerns those who believe that the quantum state is a real physical state which obeys Schrödinger's equation in all circumstances" (Pitowsky 2006, p. 232, capital in the original text), Pitowsky's point seems to be that the *big* measurement problem is a problem really *only* if we assume quantum states as *real entities*. This point makes the 'dogmatic' qualification sound less unreasonable than the case with the first 'dogma', but on what basis can we argue that it is the assumption of quantum states as 'real entities' that leads us to require from quantum mechanics a dynamical description of the measurement? And what should we exactly *mean* when we say that a state is 'a real entity'?

Let me first focus on the very problem of interpreting a physical state as a 'real-thing-out-there'. With the aid of a classical-sounding language Bub and Pitowsky depict the second 'dogma' as the extension to the quantum realm of an assumption taken to be obviously unproblematic in the realm of classical physical theories, namely that in these theories a physical state is a 'real-thing-out-there'. Moreover, they seem to assume that, for a state of a classical theory, to be a "truthmaker for propositions about the occurrence and non-occurrence of events" or a "representation" of physical reality is equivalent to be a 'real-thing-out-there'. As a matter of fact, things do not seem so straight: even if we put aside the remark that one thing is to say that a state is a real entity and quite another to say that 'it represents something in physical reality', also in classical theories the relation between a 'state' (according to a given theoretical framework) and the domain of 'real-things-out-there' – be they medium-size objects or macroscopic events or properties – is complex and far from direct.

Let us consider briefly a classical-mechanical framework. In this case according to the usual intuition, well-entrenched into the formal detailed development of any such framework, the objects the theory is about can be considered as real entities, endowed with well-defined physical properties that can be easily imagined as *possessed* properties, quite independently from any attempts on our part to check the possession of such properties on an experimental basis. In this framework, states can be conceived as *ways in which things-out-there stand*. Namely, at a time $t$ we assume a classical physical system $S$ (for simplicity, a Newtonian one-dimensional point-particle) to be in a given, conventional state represented by a pair of values for position $x$ and momentum $p$ and all remaining significant properties of $S$ depend on the values of $x$ and $p$. The set of all such points, endowed with a suitable geometric structure, is the phase space: all the physical quantities that are assumed to be relevant to $S$ (the *classical observables* for $S$) are introduced as continuous, real-valued functions on its phase space and the theory provides the formal recipe for describing the dynamics of the system. If, for whatever reasons, we are unable to specify exact values for $x$ and $p$ at a certain time, we describe the state of the system via a probability density r($x$, $p$), for which too – via the Liouville equation – a dynamics is secured. Therefore, if these are the ordinary intuition and the formal implementation of the notion of state in a classical mechanical framework, we may ask ourselves: did we somewhere need to assume that a state is – or needs to be – an *entity*? What kind of feature is such entity-language supposed to pick out exactly? And in what sense is this way of expressing the notion of state in these theories supposed to be 'representational'? The above remarks on the role of the notion of state in a classical setting, a setting in which the intuition of pre-existing entities whose properties are independent from our attempts to have access to them is unproblematic, suggest that *we need not require from states to be 'entities' in their own right*[4]. We do not at the intuitive level – in which states are not physical entities in themselves, but rather *ways of being* of physical entities – and we do not at the mathematical level – in which states are abstract, mathematical objects whose status of 'entities' is at best controversial and, in any case, in need of an engaging, Platonic-sounding argument in support.

Therefore, if the very assumption of states as real entities, in addition to its lack of clarity, appears to be unnecessary already in a non-quantum framework, it turns out to be even more dubious that such assumption is in fact assumed, explicitly or implicitly, in the quantum realm. On the contrary, the Pitowsky-Bub viewpoint not only takes this assumption to be a 'dogma', namely a statement that is endorsed uncritically, but also claims that it is *only* such an assumption that generates the 'big' measurement problem. Let us see, then, why the standard formulation of the 'big' measurement problem nowhere requires any assumption on the entity status of quantum states and why, as a consequence, the emergence of the 'big' measurement problem can be safely taken to be independent from the issue of the 'reality' of states.

In that formulation, we assume quantum mechanics to describe measurement as a special kind of interaction, such that with the coupling <measured system + measuring

---

[4] Even less so in a a classical *statistical* mechanical framework, in which a key role is played by the distinction between the macro-state and micro-state of a system, in view of the explanation of the irreversible thermodynamic behavior of macroscopic systems.

apparatus> determines a joint system whose states are supposed to evolve according to the main dynamical law of the theory, i.e. the Schrödinger equation (at least, up to the time of the measurement). Since in a measurement we are supposed to record an outcome for a physical quantity (which is well-defined for the measured system at hand), there are two possible scenarios:

(i) if the measured system' state is an eigenstate of the physical quantity to be measured, the state of the joint system will be a state in which the component referring to the measuring apparatus will be unequivocally associated to the reading of (eigen)value of the quantity pertaining the measured system;

(ii) if the measured system' state is not an eigenstate of the physical quantity to be measured, the state of the joint system will be a superposition, each component of which will be the product of measured system' state and the measuring apparatus' state, each corresponding to one of the possible (eigen)values for the physical quantity.

Now in the (ii) situation, namely when the measured system is in a superposition before the measurement takes place, the measurement problem amounts exactly to the fact that the following conditions cannot hold together:

> **C** – The wave-function associated to the state of a system is a *complete* description of the state itself, namely there can be no finer specification of the properties that the system can exhibit in the event of a measurement;
>
> **L** – The wave-function associated to the state of a system always evolves according to the Schrödinger equation;
>
> **D** – Measurements always provide have determinate outcomes, namely at the end of the measurement process the measuring apparatus is found to be in a state that indicates which among the possible values turns out to be *the* outcome of the process itself.

Even if we suppose, for the sake of the argument, to distinguish between a state and a wave function, taken as the mathematical object that 'describes' the state, still the above argument *nowhere* depends on the assumption that either the wave functions or the states are *real entities*[5].

---

[5] The point is made in especially clear terms in Maudlin 1995. A *caveat* should be made here. The above discussion on states as 'real' entities has been conducted with reference to a rather intuitive view of what it means for something to be 'real'. Namely, no specification of the issue has been attempted with the use of the conceptual resources of the so-called *ontic* vs. *epistemic* models of quantum states. For a recent paper that assesses the Pitowsky view on the issue on the background of those models, see Ben-Menahem 2020, whereas a recent evaluation of the Harrigan, Spekkens 2010 categorization and its

## 3. A 'kinematical' solution to the measurement problem?

I have argued above that the dependence of the big measurement problem on the two dogmas is controversial at best, but now let me focus on the very formulation of such problem by Bub and Pitowsky. When they say that "the 'big' measurement problem is the problem of explaining how measurements can have definite outcomes, given the unitary dynamics of the theory: it is the problem of explaining how individual measurement outcomes come about *dynamically*" (Bub, Pitowsky 2010, p. 438), they refer to a peculiar interpretation of 'dynamics'. The conceptual framework for this interpretation, apparently independent from the description of any dynamical process *per se* governing the evolution of quantum states over time, relies on two major elements:

(i) an information-theoretic view of quantum mechanics, in which the latter should be "viewed not as first and foremost a mechanical theory of waves and particles […] but as a theory about the possibilities and impossibilities of information transfer." (Clifton, Bub, Halvorson 2003, p. 1563, CBH from now on);

(ii) on the background of (i), the adoption and extension to the quantum realm of the distinction between 'kinematical' and 'dynamical', as introduced in the debate on the foundations of special relativity.

I will first outline the Bub-Pitowsky information-theoretic (IT-) view of quantum mechanics in a very sketchy way that is useful to our purpose, and after I will pass on to the analysis to the extended formulation of the pair 'kinematical'/'dynamical' that Bub-Pitowsky take to be relevant for their IT-view of quantum mechanics. This extended formulation is claimed to contribute to the (dis)solution – within the IT-view of QM – of what Bub-Pitowsky call the big measurement problem: I will argue in the section 4 that such formulation is based on a controversial interpretation of of the pair 'kinematical'/'dynamical' itself in special relativity and that, as a consequence, the 'big' measurement problem fails to be solved simply by the adoption of a Hilbert space structure with its 'kinematical' constraints. In fact, this interpretation assumes a certain kind of explanatory priority of the 'kinematical' over the 'dynamical' that is unwarranted: if we have reason to argue that this assumption fails to be convincing, then also the explanatory power of the non-Booleanity of the Hilbert space as-a-kinematical-framework can be questioned.

---

relation with the map of the different interpretation of quantum mechanics is provided in Oldofredi, Lopez 2020.

What is nowadays known as the information-theoretic view of quantum mechanics is not a monolithic view, and a detailed analysis of its possible variants is out of the scope of the present paper [6]. The IT-view of QM certainly provided a new twist to what Henderson called the 'reaxiomatisation programme in quantum mechanics' namely "a programme which aims to reaxiomatise the theory in terms of postulates which are clearer, more 'reasonable' and more physically motivated." (Henderson 2020, p. 292). In this spirit, the IT-view of QM is in general lines an attempt to ground the interpretation of quantum mechanics on principles of an informational character, principles that are taken to be "fundamental information-theoretic 'laws of nature' " (CBH 2003, p. 1562,), although the extent to which these principles should be interpreted in realistic terms, i.e. as principles that informationally constrain the very nature of physical reality, is a matter of dispute (see again Dunlap 2022 on the point). The contribution of the CBH work to the development of the IT-view was the possibility to show that, on the basis of these principles, standard Hilbert space QM could be truly *derived*:

> What CBH showed was that one can derive the basic kinematic features of a quantum-theoretic description of physical systems in the above sense from three fundamental information-theoretic constraints: (i) the impossibility of superluminal information transfer between two physical systems by performing measurements on one of them, (ii) the impossibility of perfectly broadcasting the information contained in an unknown physical state (for pure states, this amounts to "no cloning"), and (iii) the impossibility of communicating information so as to implement a certain primitive cryptographic protocol, called "bit commitment," with unconditional security. They also partly demonstrated the converse derivation, leaving open a question concerning nonlocality and bit commitment. This remaining issue has been resolved by Hans Halvorson, so we have a characterization theorem for quantum theory in terms of the three information-theoretic constraints. (Bub 2005, pp. 549-550)

To be true, the program to provide the theory with an axiomatic structure based on a restricted set of "more 'reasonable' and more physically motivated postulates" is far from new. The so-called *quantum-logical* approach to QM was developed in the Sixties exactly with that aim, on the basis of the circumstance (first recognized in von Neumann 1932 and then developed in Birkhoff, von Neumann 1936) that standard Hilbert space QM provides for the set of 'yes-no experiments' performable on a quantum system a non-classical algebraic structure – the structure **L**(*H*) of orthomodular (non distributive) lattice of projection operators acting on the Hilbert space *H* associated to the quantum system in question. As a recent review on the issue clarifies, "there […] remains the question of *why* the logic of measurement outcomes should have the very special form **L**(*H*), and never anything more general. This question entertains the idea that the formal structure of quantum mechanics may be *uniquely determined* by a small number of reasonable assumptions, together perhaps with certain manifest regularities in the observed phenomena. This possibility is already contemplated in von Neumann's *Grundlagen* (and also his later work in continuous geometry), but first becomes explicit—and programmatic—in the work of George Mackey. (Wilce 2021)"[7] Moreover, from time to time, new approaches to the foundations of QM in the last decades claimed to pay attention to this reaxiomatizing attempt. Just to mention a view that is quite popular in our times, the so-called relational approach to QM was originally an instance: in his first paper on the subject Rovelli claims "that quantum mechanics will cease to look puzzling only when we will be able to *derive* the formalism of the theory from a set of simple physical assertions («postulates», «principles») about the world. Therefore, we should not try to append a reasonable interpretation to the quantum mechanics formalism, but rather to derive the formalism from a set of experimentally motivated postulates." (Rovelli 1996, p. 1639)[8].

---
[6] For an assessment of the IT-view in this direction, one can see Dunlap 2015, 2022, Henderson 2020.

Whatever the historical antecedents of its 'reaxiomatising' attitude, the IT-view ascribes to a structure like **L**(*H*) a highly relevant role, since it is a non-Boolean structure "in which there are built-in, structural probabilistic constraints on correlations between events" (Bub, Pitowsky 2010, p. 439). This is exactly the way in which the IT-view of QM justifies a peculiar use of the kinematics/dynamics distinction:

> The structure of Hilbert space imposes kinematic (i.e., pre-dynamic) objective probabilistic constraints on events to which a quantum dynamics of matter and fields is required to conform, through its symmetries, just as the structure of Minkowski space-time imposes kinematic constraints on events to which a relativistic dynamics is required to conform. In this sense Hilbert space provides the kinematic framework for the physics of an indeterministic universe, just as Minkowski space-time provides the kinematic framework for the physics of a non-Newtonian, relativistic universe. There is no deeper explanation for the quantum phenomena of interference and entanglement than that provided by the structure of Hilbert space, just as there is no deeper explanation for the relativistic phenomena of Lorentz contraction and time dilation than that provided by the structure of Minkowski space-time (Bub, Pitowsky 2010, p. 439).

---

[7] The reference is to two highly influential works of George W. Mackey, a 1957 paper and a 1963 book (Mackey 1957, 1963).
[8] A more recent and developed instance is Höhn, Wever 2017.

But what are the presuppositions of the use of such distinction in the quantum realm?

The kinematics/dynamics distinction, to the extent to which it is supposed to play a conceptual role in the Bub-Pitowsky approach, is explicitly inspired to a distinction firstly proposed by Einstein in a short but influential text, written in 1919 for *The Times*, the distinction between *constructive* theories and *principle* theories:

> We can distinguish various kinds of theories in physics. Most of them are constructive. They attempt to build up a picture of the more complex phenomena out of the materials of a relatively simple formal scheme from which they start out. Thus the kinetic theory of gases seeks to reduce mechanical, thermal, and diffusional processes to movements of molecules—i.e., to build them up out of the hypothesis of molecular motion. When we say that we have succeeded in understanding a group of natural processes, we invariably mean that a constructive theory has been found which covers the processes in question. Along with this most important class of theories there exists a second, which I will call "principle-theories." These employ the analytic, not the synthetic, method. The elements which form their basis and starting-point are not hypothetically constructed but empirically discovered ones, general characteristics of natural processes, principles that give rise to mathematically formulated criteria which the separate processes or the theoretical represenatations of them have to satisfy. Thus the science of thermodynamics seeks by analytical means to deduce necessary conditions, which separate events have to satisfy, from the universally experienced fact that perpetual motion is impossible. The advantages of the constructive theory are completeness, adaptability, and clearness, those of the principle theory are logical perfection and security of the foundations. The theory of relativity belongs to the latter. (Einstein [1919] 1954, p. 228)

Constructive theories provide a model of phenomena that is supposed to account for such phenomena in terms of a structure which is more simple and fundamental: the canonical example is the kinetic theory, able to explain thermodynamic phenomena in terms of the particles' motion. Principle theories, on the other hand, are developed by the formulation of (empirically well-founded) generalizations, such that the theories express formal conditions that the phenomena under scrutiny are *held* to satisfy: according to Einstein, relativistic theories belong to this second class [9].

The kinematical/dynamical distinction, in the interpretation that the Bub and Pitowsky take to be relevant in the quantum realm, is actually a *variant* (and not simply an extension) of the Einstein constructive/principle distinction, developed in three steps. First, Bub and Pitowsky map the kinematical/dynamical distinction onto the Einsteinian principle/constructive distinction, by pairing *kinematical* with *principle* and *dynamical* with *constructive*. Second, the <kinematical-principle>/<dynamical-constructive> [KP/DC, from now on] distinction is then articulated by applying it entirely *within* special relativity – whereas the original Einsteinian distinction assigns special relativity theory *as a whole* to the class of principle theories. In this articulation, Bub and Pitowsky adopt the context and the language of a more recent debate on the role of the KP/DC distinction in special relativity, a debate in which two diverging views confront each other on the meaning of the physical explanation of phenomena provided by special relativity. According to one view, special relativity does its job in prescribing the structure of space-time via the specification of essentially kinematic constraints, that phenomena displacing in spacetime are held to satisfy:

> That a free particle moves in a straight line is kinematical in this reckoning since such trajectories are the geodesics associated with the flat affine structure of Minkowski space–time. […] I would say that Minkowski space–time encodes the default spatio-temporal behavior of all physical systems in a world in accordance with the laws of special relativity. Special relativity is completely agnostic about what inhabits or—to phrase it more awkwardly but in a way that may be more congenial to a relationist—carries Minkowski space–time. All the theory has to say about systems inhabiting/carrying Minkowski space–time is that their spatio-temporal behavior must be in accordance with the rules it encodes. Special relativity thus imposes the kinematical constraint that all dynamical laws must be Lorentz invariant." (Janssen 2009, pp. 27-8).

---

[9] According to a standard reading of the Einstein presentation of the distinction, only constructive theories are really *explanatory*, a reading that Einstein himself appears to suggest and endorse. Since relativistic theories are included in the class of principle theories, the above reading raises the puzzling question of what might Einstein really mean when he suggests that relativistic theories do *not* explain the phenomena they cover: an possible way out is proposed by Lange 2014. By a historical point of view, the background of the Einstein distinction is interestingly explored in Giovanelli 2020.

According to an alternative view, spacetime theories must receive a dynamical-constructive understanding in that their geometrical properties and structure must be shown to depend on the details of a (quantum) theory of matter:

> Relativistic phenomena like length contraction and time dilation are in the last analysis the result of structural properties of the quantum theory of matter. […] one is committed to the idea that Lorentz contraction is the result of a structural property of the forces responsible for the microstructure of matter. (Brown 2005, pp. vii-viii, 132).
>
> In our view, the appropriate structure is Minkowski geometry precisely because the laws of physics, including those to be appealed to in the dynamical explanation of length contraction, are Lorentz covariant […] From our perspective […] it is the Lorentz covariance of the laws that underwrites the fact that the geometry of space-time is Minkowskian" (Brown, Pooley 2006, pp. 10, 14).

On the foundations of special relativity Bub and Pitowsky side with the kinematical reading, a reading that in their view justifies the establishment of an order of relevance between kinematics and dynamics: the latter, in fact, comes first but in a sort of provisional status until a kinematical account is provided, and the main historical piece of evidence for the endorsement of this order is exactly the transition from the Lorentzian approach – a *constructive* one – to the Einsteinian approach – whose kinematical character is almost universally considered to be the reason why physics has glorified Einstein rather than Lorentz. This position leads naturally to the third and final step: namely an application of the KP/DC distinction to QM in which – so it is claimed – QM should be interpreted primarily in a principle-kinematical perspective, in which the theoretical constraints concern *information transfer*. This information-theoretic view would justify in turn the direct adoption of a Hilbert space structure as a fact, with the consequence that non-distributivity of this structure turns out to be a 'kinematical' constraint in itself:

> The information-theoretic interpretation is the proposal to take Hilbert space as the kinematic framework for the physics of an indeterministic universe, just as Minkowski space provides the kinematic framework for the physics of a non-Newtonian, relativistic universe." (Bub 2020, p. 200).

According to their supporters, this option would carry with itself for free the possibility to account for two peculiar and puzzling aspects of QM: its apparent non-locality and its apparent irreducibly probabilistic status. In fact, the non-Booleanity of the Hilbert space as-a-kinematical-framework would account *both* for the existence of entangled states – those states that generate peculiar 'non-local' correlations – and for the intrinsically non-classical probabilistic structure – since a probability theory defined over a non-Boolean structure cannot be classical:

> In special relativity, the geometry of Minkowski space imposes spatiotemporal constraints on events to which the relativistic dynamics is required to conform. In quantum mechanics, the non-Boolean projective geometry of Hilbert space imposes objective kinematic (i.e., pre-dynamic) probabilistic constraints on correlations between events to which a quantum dynamics of matter and fields is required to conform. In this non-Boolean theory, new sorts of nonlocal probabilistic correlations are possible for 'entangled' quantum states of separated systems, where the correlated events are intrinsically random, not merely apparently randomlike coin tosses." (Bub 2020, p. 201).

As a final result, therefore, in the Bub-Pitowsky information-theoretic view it is the very non-Booleanity of the Hilbert space as-a-kinematical-framework that explains "how measurements can have definite outcomes", (dis)solving thereby what *they* call the big measurement problem: if we take into account the Hilbert space structure within the IT-view of QM, since what matters in this view are only the events-as-measurement-outcomes, it follows that the structure of events-as-measurement-outcomes is best hosted by the non-commutative algebraic structure of projectors, without any need to tell a 'dynamical' or 'mechanical' (as Bub sometimes calls it) story about how such events come about.

## 4. The Bub-Pitowsky (dis)solution does not work

This proposal is problematic, though. As we just have seen, it is a certain application of the <kinematical-principle>/<dynamical-constructive> distinction to the quantum realm in the Bub-Pitowsky approach that motivates the claim according to which the very adoption of a mathematical structure like a Hilbert space explains away the (big) measurement problem. This application, however, rests on a specific view of what the <kinematical-principle>/<dynamical-constructive> distinction does in special relativity: it looks safe to say that, according to this view, it is the Minkowski geometry *qua mathematical structure* that does the job. In what is now the ordinary presentation of the kinematic vs. constructive debate concerning special relativity (the presentation we referred to above when introducing the Bub-Pitowsky proposal), a radical alternative is usually presented in terms of *explanatory priority*. Either Lorentz invariance as a physical principle explains the role of the Minkowski geometry, or the other way round. As a consequence, in each of the two camps one competitor becomes the *explanans* and the other becomes the *explanandum*: *tertium non datur*. Does the alternative really need to be so radical? Is it really well-posed in these *aut-aut* terms? In fact, in relatively recent times, plausible motivations to doubt it have been put forward. According to Pablo Acuna, for instance,

> a more nuanced, adequate and fruitful construal of the explanatory foundations of special relativity is that Lorentz invariance and Minkowski structure do not constitute two features of the theory such that one has to be explained by the other. Rather, they can be understood as *two sides of a single coin*, so there is no need to demand for an arrow of explanation connecting them. (Acuna 2016, p. 8, my emphasis)

According to the point of view advocated by Acuna, the "demand for an arrow of explanation" is far from mandatory: taking this demand to be mandatory is what turns both the kinematical (Janssen) and the dynamical (Brown) views into "overinterpretations of the connection between Lorentz invariance and Minkowski spatiotemporal structure" (Acuna 2016, p. 9). In place of a strict choice between the two views, Acuna proposes to read the Minkowski spatiotemporal structure as the 'conceptual unfolding' of the physical content expressed by the constraint coded into Lorentz invariance. There appears to be in fact a sort of mutual implication between the geometrical and the physical dimension of special relativity as a whole theory, a mutual implication that accords well also the historical process in the development of special relativity:

> what Minkowski did was not to provide for the physical grounds of the results of Einstein's (1905) paper. Minkowski's contribution is a conceptual *display* of Einstein's work, in the sense of an overt description of the spatiotemporal structure underlying the theory—a structure that Einstein had already glimpsed (Acuna 2016, p. 9, emphasis in the original).

This new reading of the kinematic/dynamic interplay in the foundations of special relativity is illustrated with the aid of a figure in which the role of the above mentioned mutual implication between Lorentz invariance and Minkowski spatiotemporal structure is suitably emphasized:

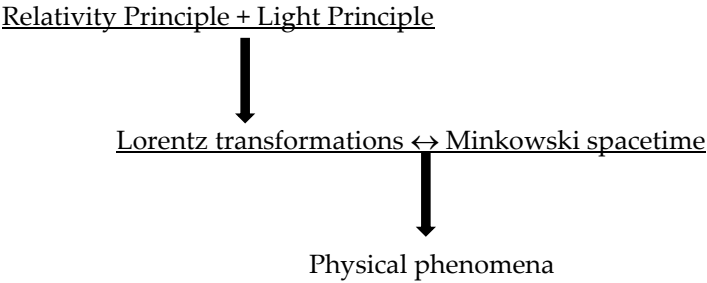

I argue that this highly plausible picture casts serious doubts on the alleged (dis)solution of the measurement problem by the IT-view of QM. In this view, as we have seen, it is the non-Booleanity of the Hilbert space as-a-kinematical-framework that is supposed to explain away the (big) measurement problem, with the alleged additional advantage of happily justifying non-classical probability theory and correlations. The possibility for this Hilbert space structure to achieve all these results *qua* kinematical framework, however, depends *exactly* on reading the confrontation between Lorentz invariance on one side and Minkowski geometry on the other in terms of an 'explanatory victory' of the latter over the former. But if the need for this explanatory asymmetry is questioned, and the mutual-implication claim is carried over the quantum domain, we may argue, first, that the non-Booleanity of the Hilbert space as-a-kinematical-framework cannot be such a cheap (dis)solution of the measurement problem and, second, a kind of mutual implication between physical principles on one side and the Hilbert space structure on the other can be imagined along the lines of the pair Lorentz invariance/Minkowski geometry: namely, the Hilbert space structure (with its formal constraints, such as the non-distributivity of the structure of projection operators) works as the 'unfolding' of physical principles that *do tell* a 'mechanical' story – in Bub's terms – about how results come about in the measurement process. The plausibility in principle of a mutual-implication account for QM raises at least two points. The first is that, unlike the case with the IT-view in which we remain essentially within the boundaries of standard Copenhagenish QM, we have to select an interpretation that is able *in principle* to tell such kind of story (that is, either Bohmian mechanics or GRW). This is totally resonant with the fact that this species of interpretations take seriously the existence of a genuine measurement problem in QM, but still this move forces us to confront with the details of these interpretive frameworks. The second is that, in the case of special relativity, the physical side – Lorentz invariance – is represented by a sort of *meta-nomological* statement, and there seems to be nothing in the quantum realm that plays a similar role at the level of the foundations of the theory.

No matter how serious this last difference should be considered, let us take into account a physical principle (referred to as 'Doctrine Q'[10]) that might play *in the quantum world* a role that is somewhat similar to the role that Lorentz invariance plays for the kinematical/dynamical distinction regarding special relativity. The Goldstein et al. 2011 formulation of this doctrine goes as follows:

**Doctrine Q** – *It is a general principle of orthodox formulations of quantum theory that measurements of physical quantities do not simply reveal pre-existing or pre-determined values, the way they do in classical theories. Instead, the particular outcome of the measurement somehow "emerges" from the dynamical interaction of the system being measured with the measuring device, so that even someone who was omniscient about the states of the system and device prior to the interaction couldn't have predicted in advance which outcome would be realized.*

---

[10] This is the terminology used by Lazarovici, Oldofredi, Esfeld 2018 to denote this principle, to which they refer in the form given by Mermin 1993, p. 803: "It is a fundamental quantum doctrine that a measurement does not, in general, reveal a preexisting value of the measured property."

In support of the claim that the mutual-implication account might be significant for my argument against the IT-view treatment of the measurement problem, let us consider then the 'Doctrine Q' in a specific interpretational setting, Bohmian mechanics, in order to see how in this case the Hilbert space mathematical machinery turns out to be a suitable theoretical environment for the 'unfolding' of the content of such a physical principle.

Standard Bohmian mechanics (BM from now on) is an observer-free formulation of non-relativistic QM that does not endorse the completeness axiom of standard QM, according to which the wave function encodes the maximal amount of information that is possible to extract concerning the state of the physical quantum systems. This stance is implemented by adding to the wave function the information concerning the system *position*: as a consequence, the wave function – in addition to satisfying the Schrödinger equation – determines the particles' motion via the especially Bohmian addition to the ordinary structure of quantum mechanics, namely the so-called guiding equation. Therefore, standard BM describes quantum particles and their trajectories in physical space and time: in doing this, BM is said to provide a space-time ontology of non-relativistic quantum mechanics, namely a class of well-specified kind of objects and properties displayed in space-time that quantum mechanics is supposed to be about (being the primary target of the theory, this space-time ontology is called *primitive ontology*). In slightly more precise terms, the main assumptions of BM are (Lazarovici et al. 2018):

*Particle configuration*: There always is a configuration of $N$ permanent point particles in the universe, characterized only by their positions $X_1, \ldots, X_N$ in three-dimensional, physical space at any time $t$.

*Guiding equation*: A wave function $\Psi$ is ascribed to the particle configuration that, at the fundamental level, is the universal wave function attributed to all the particles in the universe together.

*Schrödinger equation*: The wave function always evolves according to the Schrödinger equation.

*Typicality measure*: On the basis of the universal wave function $\Psi$, a unique 'typicality measure' can be defined in terms of the $|\Psi|^2$–density[11]. Given that typicality measure, it can then be shown that for nearly all initial conditions, the distribution of particle configurations in an ensemble of sub-systems of the universe that admit of a wave function $\psi$ of their own (known as *effective wave function*) is a $|\psi|^2$–distribution. A universe in which this distribution of the particles in sub-configurations obtains is considered to be in quantum equilibrium.

---

[11] For a proof of the uniqueness result, see Goldstein, Struyve 2007

Two consequences of this theoretical framework are especially important for our purposes. First, under the assumption of quantum equilibrium for the Bohmian universe, the Born rule for the calculation of measurement outcome statistics on sub-systems of the Bohmian universe can be *derived* (where the ordinary ψ for particular sub-systems within the universe is their effective wave function). Second, BM does not introduce intrinsic properties for the (sub)systems it is about except position, since the effective wave function that describes is provided at any time *t* by a pair ($X_t$, $ψ_t$), where $X_t$ describes the actual spatial configuration of the system. But these two points jointly illustrate a possible sense in which the Hilbert space structure is the 'unfolding' of a physical principle. According to the latter, measurements of physical quantities do not play the role of revealing pre-existing values of self-standing physical quantities, hence the treatment of measurements in the theory fulfils the Doctrine Q. On the other hand, according to the former, this is accomplished in a structure that recovers the usual Hilbert space machinery (suitably interpreted), since the Born rule cannot possibly make sense outside of that formal structure.

## 5. The *In principle underdetermination* claim

In the above pages, a major role has been played by the confrontation between the 'kinematical' virtues of the Minkowskian Einstein theory and the 'dynamical' drawbacks of the Lorentz theory, where the contrived, conspiratorial and ad hoc character of the latter was especially relevant to convince the majority about its unplausibility. In his 2005, Bub attacks Bohmian mechanics by arguing that it plays with respect to the IT-view of QM the role that the Lorentz theory played with respect to the Einstein one: this attack is based on what Henderson called a form of *in principle underdetermination* claim (Henderson 2020). In the IT-view of QM, the assumption of the relevant information-theoretic constraints is sufficient – according to the CBH theorem – to single out a Hilbert space-based theoretical structure for quantum phenomena: since Bohmian mechanics is predictively equivalent to Hilbert space QM by construction, Bub argues that a constructive theory like Bohm's theory can have no excess empirical content over a quantum theory. Just as in the case of Lorentz's theory, Bohm's theory will have to posit contingent assumptions to hide the additional mechanical structures (the hidden variables will have to remain hidden), so that *in principle*, as a matter of physical law, there *could not be* any evidence favouring the theory over quantum theory. (Bub 2005, pp. 555)

The argument, relying on the impossibility of empirically discriminating between standard QM and Bohmian mechanics (this is what *in principle undetermination* is supposed to mean), clearly suggests that the motivations underlying the adoption of Bohmian mechanics, as an alternative interpretation of standard QM, should be considered as suspicious as the motivations underlying the adoption of the Lorentz theory with respect to special-relativistic phenomena. In other words, Bohmian mechanics vis-a-vis standard QM would as contrived as Lorentz theory vis-a-vis Einstein theory *only because* the former 'can have no excess empirical content' over the latter. Moreover, in Bub's view, the same fate occurs to the GRW model: although the latter may differ in principle from standard QM in terms of empirical predictions, both Bohmian mechanics and the GRW model are taken to needlessly 'add structure' to standard QM and therefore are subject to the same argument:

> On the information-theoretic interpretation, no assumption is made about the fundamental 'stuff' of the universe. So, one might ask, what do tigers supervene on?13 In the case of Bohm's theory or the GRW theory, the answer is relatively straightforward: tigers supervene on particle configurations in the case of Bohm's theory, and on mass density or 'flashes' in the case of the GRW theory, depending on whether one adopts the GRWm version or the GRWf version. […] The solutions to the 'big' measurement problem provided by Bohm's theory and the GRW theory are dynamical and involve adding structure to quantum mechanics. There is a sense in which adding structure to the theory to solve the measurement problem dynamically—insofar as the problem arises from a failure to recognize the significance of Hilbert space as the kinematic framework for the physics of an indeterministic universe—is like Lorentz's attempt to explain relativistic length contraction dynamically, taking the Newtonian spacetime structure as the underlying kinematics and invoking the ether as an additional structure for the propagation of electromagnetic effects. In this sense, Bohm's theory and the GRW theory are 'Lorentzian' interpretations of quantum mechanics. (Bub, Pitowsky 2010, pp. 452, 454)

The argument of Bub and Pitowsky is far from convincing. As far as Bohmian mechanics is concerned, the focus on the absence of excess empirical content as the *exclusive* criterion for comparing the pairs Lorentz/Einstein and Bohmian mechanics/QM fails to distinguish between the *particular* ways in which the Lorentz theory on one side and Bohmian mechanics on the other fare concerning the issue of the empirical indistinguishability from the respective rival theory. Whereas in the Lorentz theory a form of in-principle undetectability of physical effects is introduced in a totally unconventional and instrumental way, focused as it is to the very preservation of ether as the privileged frame, the overall aim of Bohmian mechanics is to describe quantum phenomena as much as possible in line with the long and honored tradition of physics in which the theory is supposed to be about 'matter in motion'. Moreover, a major factor in the interpretation is represented by a scientifically respectable account of inaccessibility of the particles' position: although the position of every quantum system is definite at all times, we are *de facto* unable to control each such position, an uncontrollability which should look far from surprising in a quantum world anyway, and which after all resembles other forms of inaccessibility to which we are used also in a pre-quantum world, such as in statistical classical mechanics. In this respect, the absence of 'excess empirical content' might be taken to be a price that we are willing to pay if we can have in return a view of the microphysical world as a more familiar world of particles in motion. There seems to be no apriori argument why this sort of explanatory virtues should be considered less valuable than the alleged explanatory virtue of information-theoretic constraints, a virtue that manifests itself in the one and only effect of being the basis for deriving a Hilbert space structure for quantum phenomena.

Moreover, there is a further point. If the measurement problem is "the problem of explaining how individual measurement outcomes come about *dynamically*", a solution at hand is already available on the market: it is the account of the measurement problem (ordinarily understood) provided by the so-called dynamical reduction approach of Ghirardi, Rimini and Weber (GRW). In principle, the GRW approach proposes a dynamical account of the measurement interaction in terms of a *rigorous* recipe on how – in the event of a measurement of a physical quantity on a system *S* by an apparatus *A* – the (non-linearly modified) dynamics of the joint system *S+A* physically produces a definite result out of the initial, entangled state of *S+A*:

> This approach consists in accepting that the dynamical equation of the standard theory should be modified by the addition of stochastic and nonlinear terms. The nice fact is that the resulting theory is capable, on the basis of a single dynamics which is assumed to govern all natural processes, to account at the same time for all well-established facts about microscopic systems as described by the standard theory, as well as for the so-called postulate of wave packet reduction (WPR), which accompanies the interaction of a microscopic system with a measuring device." (Bassi, Ghirardi 2020).

Due to the non-linear modification of the ordinary quantum dynamics, the GRW model *does* have in principle an excess empirical content, an evidence that *might* in principle favour the GRW model over standard QM. Since it is possible to articulate an ontological reading of the GRW model, it appears highly controversial to claim that "no mechanical theory of quantum phenomena an account of measurement interaction can be acceptable." (Bub 2004, p. 241), although specifying the sense in which the GRW model can tell a 'mechanical story', according to the different, possible underlying ontologies, is a non-trivial matter (Bassi, Ghirardi 2020).

## 6. Conclusions

At least since the times of the Wittgensteinian stance displayed in the *Tractatus*, an attitude toward open foundational issues proved especially tempting: that of dissolving rather than solving the problem at stake, by showing that under certain assumptions there is simply nothing to solve, namely that it is the very status of problem need not hold for the claim under scrutiny. This is the option that Jeffrey Bub and Itamar Pitowsky developed about the infamous measurement problem in non-relativistic QM, in the framework of their information-theoretic view of quantum theory. In the present paper I have tried to show that their 'deconstructing' strategy is far from convincing: not because of the central role that the motion of information plays for the foundations of the theory – this is a totally legitimate view, among the many present in the wide market of the interpretations of quantum theory – but rather because the strategy rests entirely on the claim that an exclusively 'kinematical' reading of special relativity constitutes a good account of explanation for the phenomena covered by this theory: so good that it can be extended to QM and exploited, in order to show that the (exclusively 'kinematical' reading of the) Hilbert space structure in itself is analogously a good explanation for the quantum facts, so that we do not need any detailed, 'dynamical' or 'mechanical' account of the measurement process from which the quantum facts themselves emerge. I emphasized that if there can be grounds to question the very necessity for the explanatory priority of a kinematical account over a dynamical one (or viceversa), then there can be grounds correspondingly for rejecting what I take to be a cheap (dis)solution of the measurement problem in QM.

# References

Acuna P. 2016, "Minkowski spacetime and Lorentz invariance: The cart and the horse or two sides of a single coin?", *Studies in History and Philosophy of Modern Physics* 55, pp. 1-12.

Bacciagaluppi G. 2020, "The Role of Decoherence in Quantum Mechanics", *The Stanford Encyclopedia of Philosophy*, https://plato.stanford.edu/archives/fall2020/entries/qm-decoherence/.

Bassi A., Ghirardi G. 2020, "Collapse Theories", *The Stanford Encyclopedia of Philosophy*, <https://plato.stanford.edu/archives/sum2020/entries/qm-collapse/>.

Ben-Menahem Y. 2020, "Pitowsky's Epistemic Interpretation of Quantum Mechanics and the PBR Theorem", in M. Hemmo, O. Shenker (eds.), *Quantum, Probability, Logic*, Springer Nature Switzerland, pp. 101-124.

Birkhoff G., von Neumann J. 1936, "The Logic of Quantum Mechanics", *Annals of Mathematics* Second Series, 37, pp. 823-843.

Brown H. 2005, *Physical Relativity: SpaceTtime Structure from a Dynamical Perspective*, Oxford UniversityPress, Oxford.

Brown H., Pooley O. 2006, "Minkowski Space-Time: A Glorious Non-Entity", in D. Dieks (Ed.), *The Ontology of Spacetime*, Elsevier, Amsterdam, pp. 67–89.

Brukner C. 2017, "The Quantum Measurement Problem", in R. Bertlmann, A. Zeilinger (eds.), *Quantum Unspeakables II. Half a Century of Bell's Theorem*, Springer, pp. 95-117.

Bub J. 2005, "Quantum Mechanics is About Quantum Information", *Foundations of Physics* 35, pp. 541-560.

Bub J. 2020, " 'Two Dogmas' Redux", in M. Hemmo, O. Shenker (Eds.), *Quantum, Probability, Logic,* Springer Nature Switzerland, pp. 199-215.

Bub J., Pitowsky I. 2010, "Two Dogmas about Quantum Mechanics", in S. Saunders, J. Barrett, A. Kent, D. Wallace (eds.), *Many Worlds? Everett, Quantum Theory and Reality*, Oxford University Press, Oxford, pp. 433-459.

Clifton R., Bub J., Halvorson H. 2003, "Characterizing Quantum Theory in terms of Information-Theoretic Constraints," *Foundations of Physics* 33, pp. 1561–1591.

Dunlap L. 2015, "On the Common Structure of the Primitive Ontology Approach and the Information-Theoretic Interpretation of Quantum Theory", *Topoi* 34 359-367.

Dunlap L. 2022, "Is the Information-Theoretic Interpretation of Quantum Mechanics an Ontic Structural Realist View?", *Studies in History and Philosophy of Science* 91, pp. 41-48.

Einstein A. 1919, "What is the Theory of Relativity?", in *Ideas and Opinions*, Three Rivers Press, New York, pp. 227-232.

Giovanelli M. 2020, "Like Thermodynamics before Boltzmann.' On the Emergence of Einstein's Distinction between Constructive and Principle Theories", *Studies in History and Philosophy of Modern Physics*, 71, pp. 118-157.

Goldstein S., Struyve W. 2007, "On the Uniqueness of Quantum Equilibrium in Bohmian Mechanics", *Journal of Statistical Physics*, 128, pp. 1197–1209.

Goldstein S., Norsen T., Tausk D.V., Zanghì N. 2011, "Bell's Theorem", *Scholarpedia*, 6(10), 8378.

Harrigan, N., Spekkens, R.W. 2010, "Einstein, Incompleteness, and the Epistemic View of Quantum States", *Foundations of Physics* **40, pp.** 125–157.

Henderson L. 2020, "Quantum Reaxiomatisations and Information-Theoretic Interpretations of quantum theory", *Studies in History and Philosophy of Modern Physics* 72, pp. 292-300.

Höhn P.A., Wever C.S.P. 2017, "Quantum Theory from Questions", *Physical Review* **A95**, 012102.

Janssen M. 2009, "Drawing the line between kinematics and dynamics in special relativity, *Studies in History and Philosophy of Modern Physics*, 40, pp. 26–52.

Lange M. 2014, "Did Einstein Really Believe that Principle Theories are Explanatorily Powerless?", *Perspectives on Science*, 22, pp. 449-463.

Lazarovici D., Andrea Oldofredi A., Esfeld M. 2018, "Observables and Unobservables in Quantum Mechanics: How the No-Hidden-Variables Theorems Support the Bohmian Particle Ontology", *Entropy* 20, pp. 1-17.

Mackey G.W. 1957, "Quantum Mechanics and Hilbert Space", *The American Mathematical Monthly*, 64, pp. 45–57.

Mackey G.W. 1963, *The Mathematical Foundations of Quantum Mechanics: A Lecture-note Volume*, W.A. Benjamin, New York.

Maudlin T. 1995, "Three Measurement Problems", *Topoi* 14, pp. 7-15.

Maudlin T., 2014, "What Bell Did", Journal of Physics A: Math. Theor. 47, 424010.

Mermin, N.D. 1993, "Hidden Variables and the Two Theorems of John Bell", *Review of Modern Physics*, 65, pp. 803–815.

Oldofredi A., Lopez C. 2020, "On the Classification Between ψ-Ontic and ψ-Epistemic Ontological Models", *Foundations of Physics* 50, pp. 1315-1345.

Pitowsky I. 2006, "Quantum Mechanics as a Theory of Probability", in W. Demopoulos, I. Pitowsky (eds.), *Physical Theory and its Interpretation. Essays in Honor of Jeffrey Bub*, Springer, Dordrecht, pp. 213-240.

Rovelli C. 1996, "Relational Quantum Mechanics", *International Journal of Theoretical Physics* 35, pp. 1637–1678.

von Neumann J. 1955, *The Mathematical Foundations of Quantum Mechanics*, Princeton University Press, Princeton.

von Neumann J., 1932, *Mathematische Grundlagen der Quantenmechanik*, Springer-Verlag Berlin (English transl. *Mathematical Foundations of Quantum Mechanics*, Princeton: Princeton University Press, 1955).

Wilce, Alexander, "Quantum Logic and Probability Theory", in E.N. Zalta (ed.), *The Stanford Encyclopedia of Philosophy* (Fall 2021 Edition), <https://plato.stanford.edu/archives/fall2021/entries/qt-quantlog/>